\documentclass[12pt,preprint]{aastex}

\usepackage{apjfonts}

\slugcomment{Accepted by ApJ Letters}

\newcommand{\skipthis}[1]{}
\newcommand{\uco}[1]{\mbox{$^{#1}$CO}} 
\newcommand{\cuo}[1]{\mbox{C$^{#1}$O}} 
\newcommand{\jj}[2]{\mbox{$J = #1 \rightarrow #2$}}
\newcommand{\kms}{\hbox{km\,s$^{-1}$}}
\newcommand{\Msun}{\mbox{$M_{\sun}$}}
\newcommand{\Lsun}{\mbox{$L_{\sun}$}}
\newcommand{\mout}{\mbox{$\dot{M}_{out}$}}

\newcommand{\nnh}{N$_2$H$^{+}$}

\shortauthors{Bourke et al.}
\shorttitle{L1014 Bipolar Outflow Discovery}

\begin{document}

\title {Discovery of a Low Mass Bipolar Molecular Outflow from
L1014-IRS with the Submillimeter Array}

\author{
Tyler L.\ Bourke\altaffilmark{1}, 
Antonio Crapsi\altaffilmark{1,2},
Philip C.\ Myers\altaffilmark{1}, 
Neal J.\ Evans II\altaffilmark{3}, 
David J.\ Wilner\altaffilmark{1}, 
Tracy L.\ Huard\altaffilmark{1}, 
Jes K.\ J{\o}rgensen\altaffilmark{1}, 
Chadwick H.\ Young\altaffilmark{3}
}

\altaffiltext{1}{Harvard-Smithsonian Center for Astrophysics, 60 Garden
Street, Cambridge, MA 02138; tbourke@cfa.harvard.edu}
\altaffiltext{2}{Universit\`a degli Studi di Firenze, Dipartimento di
Astronomia e Scienza dello Spazio, Largo E.\ Fermi 5, I-50125 Firenze,
Italy}
\altaffiltext{3}{University of Texas at Austin, 1 University Station C1400,
Austin, TX 78712-0259}

\begin{abstract}
Using the Submillimeter Array we report the discovery of a compact low mass
bipolar molecular outflow from L1014-IRS and confirm its association with
the L1014 dense core at 200 pc.  Consequently, L1014-IRS is the lowest
luminosity ($L \sim 0.09 \Lsun$) and perhaps the lowest mass source known to
be driving a bipolar molecular outflow, which is one of the smallest known
in size ($\sim$500 AU), mass ($< 10^{-4}$ \Msun), and energetics (e.g.,
force $< 10^{-7}$ \Msun\ \kms\ yr$^{-1}$).
\end{abstract}


\section{Introduction}

{\it Spitzer Space Telescope} observations of the Bok globule L1014,
previously classified as starless (Parker 1988), discovered an infrared
point source with protostellar colors, L1014-IRS, located near to the
globule's dust emission peak \citep{you04}.  The infrared colors of
L1014-IRS are consistent with (i) a protostar or proto-brown-dwarf in L1014
at $\sim$200 pc (Leung, Kutner \& Mead 1982), as assumed by Young et~al.\
(2004), or (ii) an intermediate mass T Tauri star ($L \sim 16$ \Lsun)
associated with a background cloud in the Perseus arm at 2.6 kpc.  

Assuming that the distance to L1014-IRS is 200 pc, its star plus disk
luminosity, assuming isotropic emission, is $\sim$0.09 \Lsun, and the best
fit to its SED is with a protostar luminosity of 0.025 \Lsun.  
If L1014-IRS has the age of a typical Class I protostar of $\sim10^5$ yr
then it has a substellar mass of only 20-45 M$_{\rm Jup}$ (Huard et~al.\
2005).  However, no spectroscopic observations exist to determine its mass
and accretion rate, and so it is unclear whether L1014-IRS is a young
protostar still acquiring a significant fraction of its final mass, or
whether it is destined to remain substellar.

More recent observations strongly support the conclusion that L1014-IRS is
embedded within L1014 (Crapsi et~al.\ 2005; Huard et~al.\ 2005), however
the evidence is still circumstantial.  It has not been shown that L1014-IRS
has the same local velocity as L1014, nor has an outflow been detected,
despite recent searches using CO with single dish telescopes (Visser
et~al.\ 2001; Craspi et~al.\ 2005).  In this paper we present Submillimeter
Array (SMA) observations of CO \jj{2}{1}\ toward L1014-IRS, to search for a
compact outflow and to resolve the distance ambiguity.

\section{Observations}

Observations of L1014 near 230 GHz were obtained with the SMA\footnote{The
Submillimeter Array is a joint project between the Smithsonian
Astrophysical Observatory and the Academia Sinica Institute of Astronomy
and Astrophysics and is funded by the Smithsonian Institution and the
Academia Sinica.} (Ho et~al.\ 2004) on 2004 August 15 and September 5. 
Zenith opacities at 225 GHz were typically in the range 0.1-0.2.  The
observations utilized both 2 GHz wide receiver sidebands, separated by 10
GHz.  The SMA correlator was configured with high spectral resolution bands
of 512 channels over 104 MHz for the \uco{12}, \uco{13}\ and \cuo{18}\
\jj{2}{1}\ lines, providing a channel spacing of 0.26 \kms, with a lower
resolution of 3.25 MHz/channel over the remainder of each sideband.
Observations of L1014 were interleaved with the quasars BL Lac and
J2013+370 for gain calibration.  The data were edited and calibrated using
the MIR software package adapted for the SMA.  Saturn and Uranus were used
for passband and flux calibration, respectively.  The flux of BL Lac
measured on the two days agrees to within 20\%, which we take to be the
uncertainty in the absolute flux scale.  Mapping was performed with the
MIRIAD package, resulting in an angular resolution of 1\farcs2 $\times$
1\farcs0 (natural weighting).  The rms sensitivity was $\sim$2 mJy for the
continuum, using both sidebands (avoiding the band containing the CO line),
and $\sim$0.1 Jy beam$^{-1}$ per channel for the line data.  The primary
FWHM beam of the SMA is $\sim$55$''$ at these frequencies.  

\section{L1014-IRS: A Protostar at 2.6 kpc?}
\label{sec-far}

The systemic velocity of L1014 is 4.2 \kms\ (Crapsi et~al.\ 2005).  Behind
L1014 lies another cloud with a velocity near $-$40 \kms\ associated with
the Perseus arm at 2.6 kpc (Young et~al.\ 2004; Crapsi et~al.\ 2005).
Emission from the \jj{1}{0}\ lines of \cuo{18}\ and \nnh, and the \jj{2}{1}
line of CS, tracing gas density 10$^{3-5}$ cm$^{-3}$, was not detected at
these velocities in a $\sim$50\arcsec\ beam by Crapsi et~al.\ (2005).  No
CO emission is detected in the SMA data over the velocity range
[$-$35,$-$45] \kms, implying that no compact CO emission nor outflow is
associated with this velocity component to the level of our sensitivity.

A tentative detection of continuum emission of 7 $\pm$ 3 mJy was obtained
at the position of L1014-IRS (offset by 4\farcs8, 0\farcs1 from the phase
center) with observations made in the compact configuration, but was not
confirmed by the extended array observations.  By combining the two data
sets and only selecting baseline lengths $<$ 50 k$\lambda$ a similar
tentative detection (5 $\pm$ 2 mJy) is obtained.  \cite{you04} predict a
flux of $\sim$1-2 mJy at 230 GHz for the central object (star+disk) of
L1014, which is consistent with our results.  Recent BIMA imaging at
$\sim$95 GHz with a $15-20''$ synthesised beam shows weak slightly extended
emission (Lai et~al.\ in preparation), but no compact component.
\cite{you04} predict that the emission from a background protostar ($L \sim
16 \Lsun$) would be $\sim$ 0.6 Jy, which is clearly ruled out by the SMA
observations, and together with the lack of
line emission at $-$40 \kms\ (\S \ref{sec-far}), rules out the possibility
that L1014-IRS is a distant young star.

\section{CO Outflow Associated with L1014-IRS}
\label{sec-outflow}

\begin{figure*}
\centering
\includegraphics[angle=270,width=6.5in]{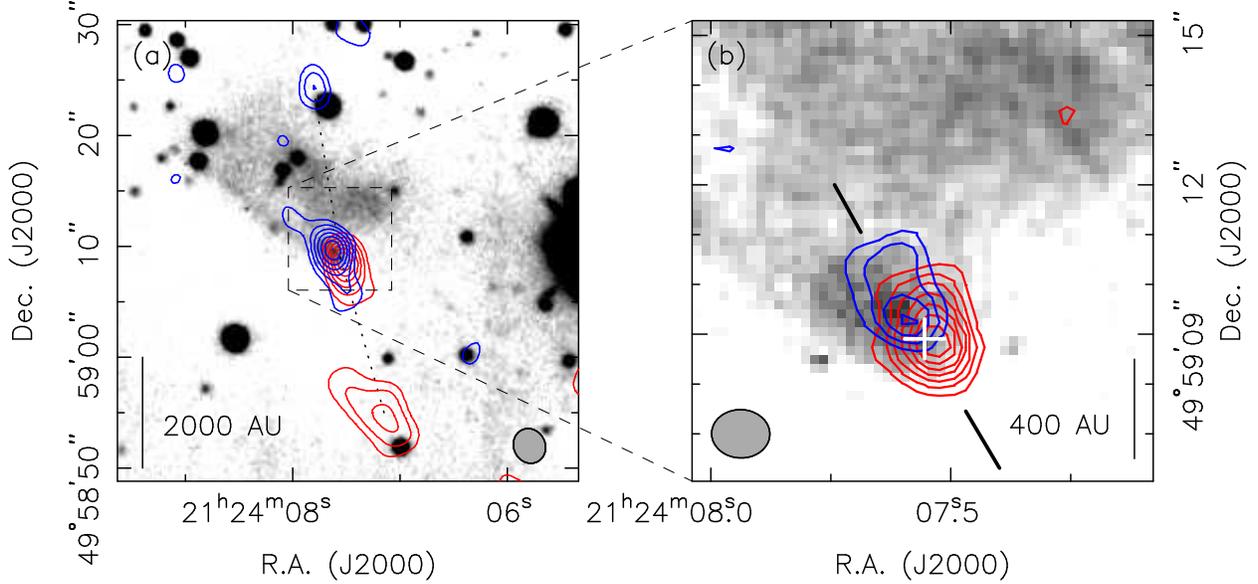}
\caption{Integrated intensity maps of CO \jj{2}{1}\ emission toward L1014,
overlaid on a H band image (1.6 \micron) from Huard et~al.\ 2005.  
{\it (a):} Blue (red) contours represent blue (red) shifted emission
between velocities (LSR) of 2.0 and 3.8 \kms\ (4.9 and 7.0 \kms).  The
images were made with natural weighting, and have been tapered with a
3\arcsec\ Gaussian.  The contours are 2, 3, 4, ... $\times$ the rms of 0.16
Jy/beam (0.25 Jy/beam) for the blue (red) shifted emission.  The dotted
lines are drawn from L1014-IRS to the peaks of the blue and red shifted
emission that are spatially offset from L1014-IRS.  The beam of 3\farcs19
$\times$ 2\farcs84, P.A. 24\fdg7 is shown at lower right.  
{\it (b):} As for (a), but with integration over 2.2 and 3.4 \kms\ (blue)
and 5.2 and 7.5 \kms (red).
The images were made with Briggs' robust parameter of 0.  The contours are
2, 3, 4, ... $\times$ the rms of 0.08 Jy/beam (0.12 Jy/beam) for the blue
(red) shifted emission.  The position and uncertainty of
L1014-IRS is indicated with the white cross, and the beam of 1\farcs17
$\times$ 0\farcs97, P.A. 88\fdg1 is shown at lower left.  The
position-velocity cut shown in Fig.~\ref{fig-pv} which passes through
L1014-IRS is indicated by the two short black lines either side of the
outflow with P.A. $\sim 30\degr$.  \label{fig-maps}} 
\end{figure*}

\begin{figure}
\centering
\includegraphics[angle=270,width=6in]{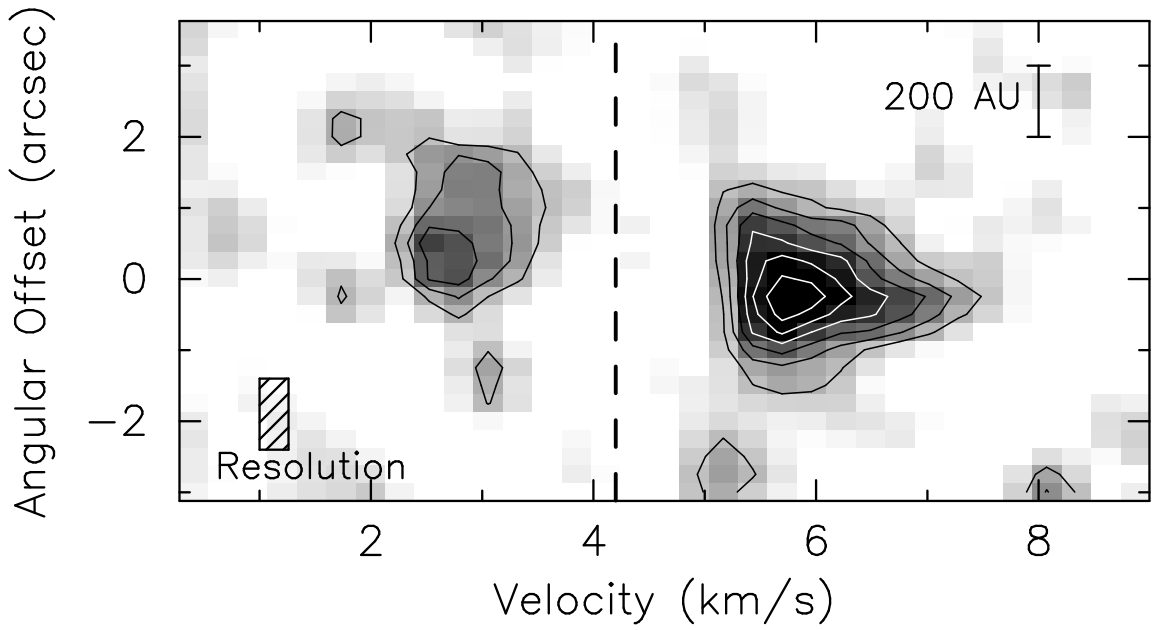}
\caption{Position-Velocity (P-V) diagram for CO \jj{2}{1}\ emission.  
The direction of the cut is indicated on Fig.~\ref{fig-maps}(b), and
offsets are relative to L1014-IRS.
The cloud systemic velocity marked by the thick dashed line.  The contours
are 2, 3, 4, ... $\times$ the rms of 0.08 Jy/beam.
\label{fig-pv}}
\end{figure}

CO \jj{2}{1}\ (hereafter referred to as CO) emission was detected at
velocities blue-shifted and red-shifted with respect to the systemic
velocity of L1014 of 4.2 \kms.  No emission was detected at the systemic
velocity.  Figure~\ref{fig-maps} presents an overview of the CO emission
integrated over the blue- and red-shifted velocities, and Fig.~\ref{fig-pv}
presents a position-velocity (P-V) diagram along the main axis of the
compact outflow (indicated on Fig.~\ref{fig-maps}(b)).  The CO emission is
clearly offset from the position of L1014-IRS, in a pattern typical of
bipolar molecular outflows, with the blue-shifted emission aligned with the
near-infrared scattered light nebula (Huard et~al.\ 2005).  Two
representations of the data are shown.  In Fig.~\ref{fig-maps}(a) the data
have been tapered with a 3\arcsec\ Gaussian in order to improve the
sensitivity to the fainter emission seen $\sim20\arcsec$ north and south of
the main outflow.  The red-shifted emission to the south is clearly visible
even at the highest angular resolution.  These faint extended emission
features are approximately equidistant from the position of L1014-IRS and
may be associated with the outflow, perhaps indicating episodic emission.
Fig.~\ref{fig-maps}(b) and Fig.~\ref{fig-pv} show that the inner bipolar
emission is very compact:  the peaks have projected separation of $\sim$200
AU centered on L1014-IRS, the lowest contours are separated by at most
$\sim$700 AU, and the size of each lobe is only $\sim$540 AU (0.0026 pc) in
length.

No emission is detected at the cloud systemic velocity, and the peak
emission in both the blue and red lobes are offset by similar amounts from
the systemic velocity (Fig.~\ref{fig-pv}).  The red-shifted emission is
clearly brighter than the blue-shifted emission by about a factor of two.
The brightness asymmetry might arise because L1014-IRS is offset from 
the density peak, with the red-shifted outflow propagating into denser
material resulting in a larger swept up mass, compared to the blue-shifted
emission.  The core's column density peak lies $\sim10\arcsec$ to the south
(red-shifted side) of L1014-IRS (Huard et~al.\ 2005), supporting this
conjecture.  Alternatively, the difference could be due to radiative
transfer effects implying a moderate CO opacity (Bally \& Lada 1983).

\subsection{Molecular Outflow Mass, Kinematics and Dynamics}

\begin{figure} 
\centering
\includegraphics[width=5in]{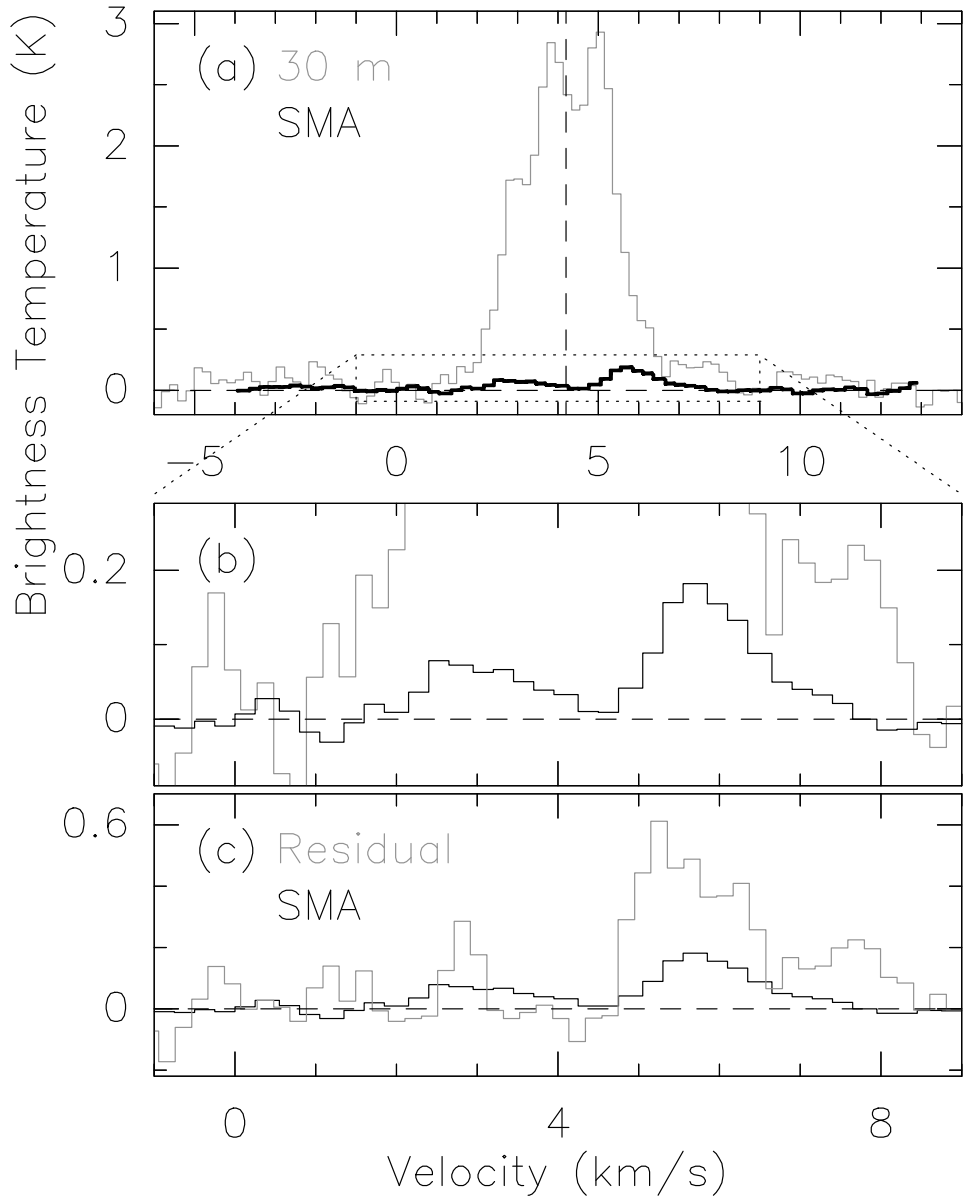}
\caption{Comparison of the IRAM 30-m and SMA CO \jj{2}{1}\ observations of
L1014.  Panel (a) shows the 30-m (11$''$ beam) and SMA spectra (convolved
with a 4$''$ beam and scaled by the ratio of the 30-m and SMA beam areas)
at the position of L1014-IRS.  Panel (b) shows an enlarged version of a
portion of panel (a) for clarity.  Panel (c) shows again the SMA spectrum,
and the difference (``Residual'') between the 30-m center spectrum and a
mean spectrum made by combining spectra at 11\arcsec\ offset positions in a
ring around the center.
\label{fig-compare}} \end{figure}

In order to derive the outflow properties we only consider the compact
outflow shown in Fig.~\ref{fig-maps}(b).  The nature of the ``extended
outflow'' component is unclear, and requires confirmation, perhaps by
combining the SMA data with single-dish data to recover extended missing
flux.  Observations with the IRAM 30-m telescope of the L1014 cloud in CO
\jj{2}{1}\ have recently been made by us.  A full analysis of these data
are beyond the scope of this paper.  However, we can investigate
qualitatively the missing flux question.  Figure~\ref{fig-compare} presents
a comparison of the SMA and 30-m data.  The SMA data have been convolved
with a 4\arcsec\ beam (i.e., the approximate area of the outflow), and
scaled by the ratio of the beam areas, to approximate the outflow emission
detectable by the 30-m in equivalent brightness temperature units.
Figures~\ref{fig-compare}(a) and (b) show that no emission and so no
missing flux is present at velocities greater than that detected by the
SMA.  Figure~\ref{fig-compare}(c) indicates that no outflow emission is
seen at positions more than $\sim 11\arcsec$ from L1014-IRS in the 30-m
data and so no bright large scale outflow is present.  In fact, there
appears to be residual emission (cloud plus outflow) detected with the 30-m
at the position of L1014-IRS some of which could be associated with the SMA
outflow.  This simple analysis suggests that any missing flux is limited in
velocity and extent, and is less than a factor of 3
(Fig.~\ref{fig-compare}(c)).

The outflow properties are calculated in the standard manner (e.g., Cabrit
\& Bertout 1990).   We assume a value for the excitation temperature of 20
K, but values in the range 10--50 K modify the outflow calculations only
slightly, less than a factor of two.  Two methods are typically used to
calculate upper and lower limits to the outflow mass.  For the lower limit
the emission is assumed to be optically thin, the mass is calculated at
each velocity channel and position where outflow emission is observed, and
then summed.  In the case of the SMA observations we are also missing some
emission that has been filtered by the interferometer sampling, and so
these values are strict lower limits.  However, Fig.~\ref{fig-compare}
suggests that the SMA observations do not suffer greatly from this problem.
For the upper limits, a correction for line opacity is required, but we
cannot estimate the opacity from our observations as we do not detect
\uco{13}\ \jj{2}{1}, and the sensitivity of the line observations is
insufficient to place any meaningful limits on the ratio of \uco{13}\ to
\uco{12}\ emission.  Thus, we derive only a lower limit to the outflow mass
of $M_{flow} \sim 1.4 \times 10^{-5}$ \Msun, but note that, if we correct
for optical depth using typical values of 2-5 (Levreault 1988), then the
upper limits to outflow mass and related properties will be correspondingly
greater than our lower limits by similar factors.  The mass estimate is
consistent with the upper limit of $2 \times 10^{-3}$ \Msun\ estimated by
Crapsi et~al.\ (2005) from their single-dish data.  

For properties relying on a knowledge of the outflow velocity (Momentum
$P$, Energy $E$, Mechanical Luminosity $L_m$, and Force $F_{obs}$), lower
limits are determined by multiplying the mass in each velocity channel by
the relative flow velocity (to the appropriate power) of that channel
$v_{flow} = \mid V_{flow} - V_{lsr} \mid$ where $V_{flow}$ is the observed
flow velocity and $V_{lsr}$ is the systemic velocity.  Upper limits are
found by assuming that the outflowing gas is moving at the maximum observed
velocity $V_m$, rather than the relative flow velocity of the channel,
corrected for the outflow inclination.  From modeling of the infrared
scattered light nebula Huard et~al.\ (2005) determine a semi-opening angle
of $\theta \geq50\degr$, and an inclination angle $i > 60\degr$.  The
correction factor $1/\cos(i-\theta)$ for inclination effects is then
negligible (for $i \leq 80\degr$).  The overlap between the blue and red
lobes seen in Fig.~\ref{fig-maps}(b) suggests that $i$ is closer to
60\degr\ than $>75\degr$ (Cabrit \& Bertout (1990)).  Regardless of the
true value for $i$, if $20\degr < i < 80\degr$ then the correction for
inclination is negligible due to the large observed opening angle.
Therefore, for the upper limit values we use the observed maximum flow
velocity of 3.6 \kms\ for $V_m$, assuming no correction is needed for
inclination.  In this way we find $P = 2.4-5.0 \times 10^{-5}$ \Msun\ \kms;
$E = 0.24-1.6 \times 10^{-4}$ \Msun\ km$^{2}$ s$^{-2}$; $L_m = 0.34-2.8
\times 10^{-5}$ \Lsun; and $F_{obs} = 3.4-7.1 \times 10^{-8}$ \Msun\ \kms\
yr$^{-1}$.  If correction factors for opacity of $\sim3$ and missing flux
of $\sim3$ (Fig.~\ref{fig-compare}) are assumed, then the maximum values
for $P, E, L_m$ and $F_{obs}$ will increase by about an order of magnitude.  

The outflow mass loss rate \mout\ can be estimated directly from the mass
and age $t_{flow}$.  Typically $t_{flow} \sim t_d$ (the dynamical time) is
assumed, which may not be reasonable (Parker et~al.\ 1991); $t_d$ is a
lower limit to the outflow age.  Assuming $t_{flow} \sim t_d = 700$ yr
(assuming a lobe size of 540 AU, see \S \ref{sec-outflow}), and allowing as
above for an increase in the mass by $\sim10$ due to opacity and missing
flux corrections, we find an upper limit of \mout\ $\sim 2 \times 10^{-7}$
\Msun\ yr$^{-1}$.  If instead $t_{flow} \sim 10t_d$ with no correction
applied to the mass estimate we obtain \mout\ $\sim 2 \times 10^{-9}$
\Msun\ yr$^{-1}$.

\subsection{Discussion}

The values calculated above show that the outflow is of low mass, and
correspondingly weak in its momentum, energy, mechanical luminosity, and
force, when compared to other outflows (Bontemps et~al.\ 1996; Wu et~al.\
2004).  In fact, L1014-IRS is the lowest luminosity source to date with a
detectable bipolar molecular outflow, of 292 outflows with well determined
bolometric luminosity $L_{bol}$ (Wu et~al.\ 2004).  Our data suggests that
L1014-IRS is not edge on, and so its $L_{bol}$ estimate is reasonable.
Even allowing for an upward revision of the outflow parameters due to
uncertainties in the CO opacity and missing flux, the outflow mass is at
least an order of magnitude lower than any other outflow; the lowest
previously reported has $M \sim 10^{-3}$ \Msun\ (B1-IRS; Wu et~al.\ 2004).
The derived values of $L_{bol}$, $M$, $L_m$, and $F_{obs}$ are consistent
with extrapolation of the relations among these quantities found for other
outflows (e.g., Wu et~al.\ 2004).  L1014-IRS appears to be a weaker version
of a typical outflow.  The low values might suggest that the emission is in
fact due to bound motions and not outflow emission, so using a size of 540
AU and velocity of 3.6 \kms\ we estimate an interior mass of $\sim$4 \Msun\
is required for bound motions, significantly larger than the core mass
within the same radius ($<$0.5 \Msun; Huard et~al.\ 2005).

For comparison, the outflow from the low luminosity Class 0 protostar IRAM
04191+1522 ($L \sim 0.15\Lsun$ cf. 0.09\Lsun\ for L1014-IRS; Andre, Motte
\& Bacmann 1999) is large ($> 14000$ AU) with $M \geq$ 0.03
\Msun\ (Lee et~al.\ 2002 from BIMA data) and $F_{obs} \sim 2 \times 10^5$
\Msun\ \kms\ yr$^{-1}$ (Andre et~al.\ 1999).  Unlike L1014-IRS, the CO
outflow from IRAM 04191 is well detected with the IRAM 30-m over a larger
velocity range, up to 10 \kms\ from the line center.  The implied accretion
rate, which follows directly from $F_{obs}$, is over two orders of
magnitude greater for IRAM 04191.  In addition, IRAM 04191 shows clear
evidence for infall in mm transitions of CS and is bright and extended in
mm molecular lines such as \nnh\ (Belloche et~al.\ 2002; Crapsi et~al.\
2005).  So although L1014-IRS and IRAM 04191 have similar low luminosities
and are embedded in dense cores, the properties of their outflows and cores
show significant differences, which may signify different pathways to
formation, and different end results.  Its outflow properties suggest that
L1014-IRS has a low accretion rate for an embedded source, and we can
speculate that it is either (i) close to finishing its main accretion
phase, or (ii) its accretion rate is intrinsically low.  Either scenario
implies that L1014-IRS will remain a very low mass object, and it may be
the first example of an embedded proto-brown dwarf.  Further information,
such as a near-infrared spectral classification, is needed to address its
evolutionary state and future.  

If L1014-IRS is indeed a proto-brown-dwarf, then our results suggest that
BDs can form in a broadly similar manner to low mass stars (cf., the
recent detection of outflowing gas from the young brown dwarf $\rho$ Oph
102 by Whelan et~al.\ (2005)), but that there might be significant
differences in the details, by comparison with IRAM 04191.  Alternatively
the differences between L1014-IRS and IRAM 04191 might simply reflect
differences in the mass and evolutionary state from one low-luminosity
object to the next.

\section{Summary}

We have used the SMA to search for CO \jj{2}{1}\ outflow emission from
L1014-IRS.  The results are as follows: 
\vspace*{-6pt}
\begin{enumerate}
\itemsep-3pt
\item A low velocity ($<$ 4 \kms) compact bipolar CO outflow has been
discovered centered on L1014-IRS, which directly associates L1014-IRS with
the globule L1014 at $\sim$200 pc and confirms its low luminosity
($\sim$0.09 \Lsun).
\item L1014-IRS is the lowest luminosity and perhaps the
lowest mass source known to be driving a bipolar molecular outflow.
\item The CO outflow is one of the smallest known in size
($\sim$500 AU), mass ($< 10^{-4}$ \Msun), and energetics (e.g., $F_{obs} <
10^{-7}$ \Msun\ \kms\ yr$^{-1}$).  These values are consistent with
the trends observed between them for large samples. 
\end{enumerate}

\vspace*{-12pt}
\acknowledgments
T.~L.~B. acknowledges support from the SMA Fellowship
Program.  The work of A.~C.\ was supported by a Smithsonian Predoctoral
Fellowship.  P.~C.~M.
acknowledges support from NASA Origins of Solar Systems Program Grant NAG
5-13050.  We thank Mario Tafalla for obtaining the 30-m data.  We extend
special thanks to those of Hawai'ian ancestry on whose sacred mountain we
are privileged to be guests.

\clearpage

\clearpage

\clearpage

\end{document}